\documentclass[A4paper,12pt,twocolumn]{article}
\usepackage[dvips]{graphicx,epsfig}
\begin{document}
\title{A study on the size of snow particles in powder snow avalanches}
\author{Marie CLEMENT-RASTELLO\thanks{This work was supported by the Pole Grenoblois sur les Risques Naturels}\\
Cemagref, u.r. etna, 2 rue de la papeterie, BP 76,\\ 38402 Saint Martin d'H\`eres France}
\date{}

\maketitle
\abstract{In this work, we study the size of the particles involved in a powder snow avalanche phenomenon. To determine these sizes, we study 
all the phenomena the particles have to face before arriving in the ``body'' of the avalanche.
We study the boundary layer which is at the bottom of the avalanche. 
We determine, with the help of experimental data, the range of size of the particles that can be entrained by the avalanche. We then examine 
the possibility for these particles to reach the top of the boundary layer, and so to 
take part in the avalanche. Our final result is that the more frequent particles suspended in a powder snow avalanche have a size 
lower than $200\mu m$.}

\section{Introduction}
Powder snow avalanches are very destructive and very ill-known phenomena. In this article we try to study them with a ``microscopic'' scale 
point of view. This kind of approach has been recently used on experimental sites where people try to know not only the macroscopic 
but also the inner properties such as pressure, velocity, etc. of avalanches (see Dent and others (1998), 
Nishimura and others (1997), Qiu and others (1997)). 
The purpose of this study is to know a bit better the size of the snow 
particles that are encountered in a powder snow avalanche. This has a real importance for someone who wants to study the interaction between 
particles and the turbulence of the flow. Indeed it is important to know the size of the 
particles that are involved in the avalanche so as to compare it with the characteristic scales of the turbulent flow, to determine the 
difference of velocity between air and particles. The study presented 
here is a theoretical study in which we try to find out which sizes of particles are encountered more frequently in a powder snow avalanche.\\
The way we proceed is the following: we study separately each of the mechanisms that are encountered by a snow particle before reaching the body 
of the avalanche. For each of them, we look 
at the range of sizes of the concerned particles. At the end we compare the different ranges found in each case so as to find the range that we must 
consider for a powder snow avalanche. The study is two-dimensional and deals neither with the front part of the avalanche nor with its tail but 
with its body. 
A pure powder snow avalanche is studied going its way on the snow cover. The different phenomena that are considered are: the size of the particles in the snow cover, the picking up 
of the particles from the snow cover by air friction and by collisions, and finally the proportion of each type of particles that manage to reach the 
top of the boundary layer, which is at the bottom of the avalanche, and so that take part to the body of the avalanche. For this study, we need 
experimental data. We chose to use the results obtained by Nishimura and others (1997) while making measurements in an effective avalanche.    

\section{The particles of the snow cover}
A first range for the size of the particles is given by the particles which are present in the snow cover. This range of size is not 
a fixed parameter, it depends on many parameters such 
as temperature, wind velocity, etc. during and since the snow fall. Mellor (1964) studied the size of the particles that can be found 
in snow layers. We use his results concerning recent snow covers. From these we conclude that the size of 
the more frequent particles is smaller than $1-2~mm$. 

\section{The pick up of the particles from the ground}

\subsection{Mechanisms \label{coll}}
A particle on the ground can be set into motion either due to air friction or 
to the impact of another particle which is projected into the ground.\\
Here, the aim is to obtain the size of the potentially mobilized snow grains.
Let us first look at the collisions. What we show here is that the collisions 
between particles and the snow cover can increase the range of size with small radius 
but not really with large radius. Indeed let us first suppose the ``best case'' of energy transmission. One particle of radius $a_{c}$ arrives 
with a velocity $u_{c}$. If it collides elastically with a particle of the snow cover of radius $a_{l}$ it gives all its energy to this particle. In this case the 
leaving particle will have a velocity 
\begin{equation} 
u_{l} =(\frac{a_{c}}{a_{l}})^{\frac{3}{2}}u_{c}.
\end{equation}
One can see that even in this ideal case to have a significant velocity the leaving particle cannot have 
a size much larger than the incoming one. In reality it seems clear that the transmission of energy is not so perfect and that we do not have all 
the energy transferred from one particle to another. So we have 
\begin{equation} 
u_{l} \le (\frac{a_{c}}{a_{l}})^{\frac{3}{2}}u_{c}
\end{equation}
and our previous conclusion remains valid.
On the other hand the collision phenomenon does not give any restriction on the lower limit of the sizes of the particles. So when we will 
compare the results of the picking-up by collisions and of the picking-up by air friction, because of the collisions we will lose the information on 
the lower limit. For the upper limit the collisions will have no influence.
Let us now look at the range we obtain from 
pick up by air friction.

\subsection{Threshold velocity for particle pick-up \label{se}}
\begin{figure}[htbp]
\begin{center}
\epsfig{file=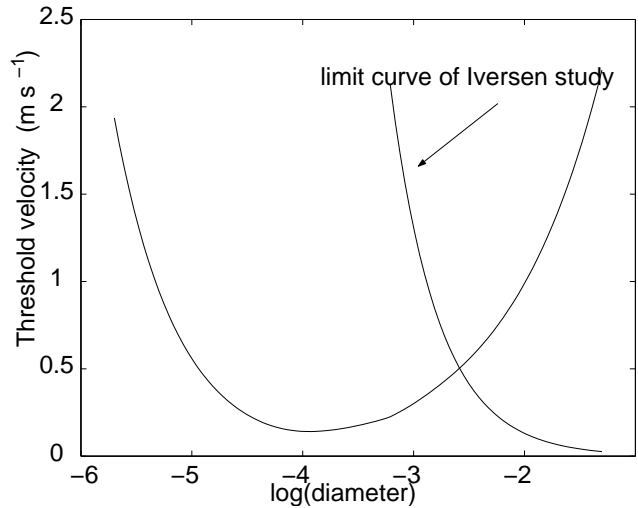,width=8.5cm}
\caption{Threshold friction velocity versus diameter of the particles \label{u}}
\end{center}
\end{figure}

In order to determine the radius of the particles which are entrained from the snow cover by the 
friction of the avalanche, we use results that have been established for sediment transport by the wind 
and that are also used in snow transport by the wind. Here we use the results given by Iversen and White (1982) 
(see Fig.~\ref{u}). 
To collect their data they placed samples of material on a wind tunnel floor and determined optically the threshold friction 
velocity ($u_{t}^{*}$) for each of them. The particles they used were of different kind and had different shapes. From their 
experimental results they deduced the following law:
\begin{equation}
A=
\left\lbrace
\begin{array}{l}
0.129\sqrt{\frac{1+\frac{6*10^{-7}}{\overline{\rho}_{p} g d^{2.5}}}{1.928 B^{0.092}-1}}, \ \ \ \ 0.03\le B\le 10\\
0.120 \sqrt{1+\frac{6*10^{-7}}{\overline{\rho}_{p} g d^{2.5}}}\\
 *(1-0.0858 e^{(-0.0617 (B-10))}), \ \ \ 10\le B
\end{array}
\right.
\end{equation}
with $A=u^{*}_{t}\sqrt{\frac{\overline{\rho}}{\overline{\rho}_{p} g d}}$ and $B=\frac{u^{*}_{t} d}{\nu}$, where $d$ is the diameter of a particle, 
$A$ is the ratio of the 
threshold friction velocity to the shallow water velocity of a density current and $B$ is a Reynolds number.\\
Iversen and White give a lower limit ($B=0.03$) but no upper limit for their law. The lower limit is outside the range of our study so does not 
limit us. Concerning a possible upper limit we plot on the figure the curve which separates the region in which Iversen and White made measurements 
and that where they did not make any. 
They say that for particles larger than $40~\mu m$ the above empirical law is generally within $5\%$ of the experimental values. 
For smaller particles it is more difficult to perform the experiments and so the precision is lower. 
This is of no importance here because due to the collisions, information on the lower limit is lost. 
As usually done in snow transport studies we use this law here. One restriction would be that in their experiments Iversen and White had no 
adhesion between the particles, as we can have with snow. We suppose that for powder snow such bounds are not predominant. Another restriction is 
that the experiments have been performed each time with one given size of particles, so that the different phenomena that can 
occur when there are both small and large particles are not taken into account. A further major restriction is that Iversen and White's results 
were established in clean air on a horizontal floor, whereas in our case the air is loaded with particles and the ground can be inclined. 
From a recent study by Bintanja (1998)
 it is inferred that the presence of particles modifies some characteristics of the boundary layer. 
However there are insufficient results that could be taken into account in a modified boundary layer model. So we use a clear air 
velocity profile essentially for lack of better knowledge. 

\subsection{Determination of the friction velocity in a powder snow avalanche}
Now that a relation between the threshold friction velocity and the diameter of the potentially mobilized 
particles is available, the friction 
velocity in a powder snow avalanche must be determined. 
The avalanche studied by Nishimura and others was a ``mixed avalanche'': 
there was a dense part and above it a powder snow part.
We suppose that the internal structure of the powder snow avalanche is not modified by the presence of the dense 
avalanche and use the results of the measurements for the pure powder snow. The velocity of the front part of the 
avalanche was from $40~m.s^{-1}$ to $60~m.s^{-1}$.
At a given point, the static pressure was measured and compared to the static pressure 
measured somewhere outside the influence of the flow. From this the velocity at the measuring point was measured and for 
the part of the avalanche behind the front a value of $10~m.s^{-1}$ was found.
The point where the measurements were taken was located less than seventy centimeters above the bottom 
of the powder snow part; this is in the boundary layer of the powder snow avalanche. 
Let us suppose that the boundary layer is of the order of one meter thick. Later on we will see that the exact value of the size is 
of no importance. Thus,
if $z\le 25~cm$ ($z\le 0.25\delta$ with $\delta =1~m$ the thickness of the boundary layer), the value of the friction velocity can be 
directly estimated from the logarithmic law which governs the velocity in this region~:  
\begin{equation} 
u=\frac{u^{*}}{K} ln(\frac{z}{z_{0}})
\end{equation}
where $K=0.41$ is the von K\`arm\`an constant, $z_{0}$ the roughness height, 
$z_{0}=\frac{C_{0} (u^{*})^{2}}{2g}$, with $C_{0} =0.021$ (see Owen (1964) and Rasmussen and Mikkelsen (1991)). 
\begin{figure}[htbp]
\begin{center}
\epsfig{file=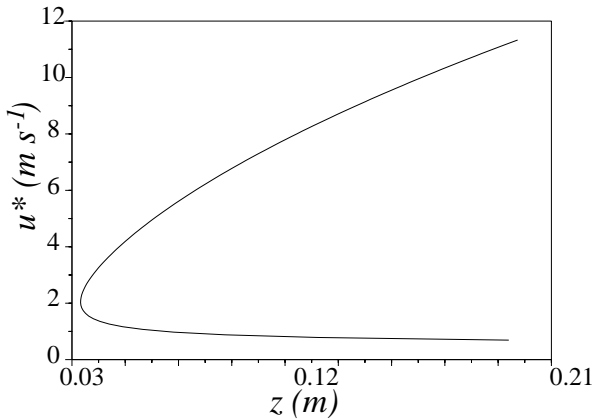,width=8.5cm}
\caption{Determination of $u^{*}$ \label{detu}}
\end{center}
\end{figure}
There are, theoretically two 
possible values for $u^{*}$ (see Fig.~\ref{detu}).\\
In 
comparison with the values of $u^{*}$ found in snow 
transport by wind, with velocities comparable to those of avalanches, it appears 
that the lower value is the physically relevant value. We deduce from Fig.~\ref{detu} that in this region $ max(u^{*})\approx 1~m.s^{-1}$.\\
In the upper region ($25~cm\le z\le 70~cm$) the boundary layer profile relating velocity and height is unknown. 
If the measurement has been done in this region, then $u(z\epsilon [0; 0.25])\le 10~m.s^{-1}$. 
So with the formula relating $u$ and $u^{*}$ in the logarithmic region, one necessarily has $u^{*}\le 1~m.s^{-1}$. 
Thus here again $max(u^{*})= 1~m.s^{-1}$.\\ 
Let us now suppose that $\delta\ne 1~m$. With the same reasoning as before one can show that if $u$ has been measured in the upper region 
the deduced value of $u^{*}$ is lower than it would be if $u$ had been measured in the logarithmic region. Because in the logarithmic 
region $u^{*}$ is decreasing with z (see Fig.~\ref{detu}), for any $\delta$, for $z\le 0.25\delta$ the value of $max(u^{*})$ remains $1~m.s^{-1}$.
Thus the size of the boundary layer does not change anything in this estimate.\\
Two things should be improved in the future~: first a better determination of $u^{*}$ with more field data, and second to account for the presence 
of the particles in the boundary layer in the determination of the velocity profile $u(z)$.

\subsection{Size of the entrained particles}
We are now able to determine the maximum range of sizes of the particles which can be picked up from the snow cover by air friction~: 
it is $[5*10^{-6}~m; 10^{-2}~m]$ (see Fig.~\ref{u}). To determine these values of the size ($d$) of the particles, formulas 
given by Iversen and White were used (see section~\ref{se}). The upper value of $d$ still lies in the domain of validity of their 
formulas but is outside the region in which they performed the experiments. Because they did not put any upper limitation 
to their formula and because our range is not far from the range they studied experimentally we think their formula is applicable for the upper 
value of $d$.\\
The upper limit of this interval is not changed by the collisions of 
particles with the snow layer, but the lower limit can be changed by the impacts (see section~\ref{coll}). Considering both air friction and 
collisions, the size of the mobilized particles is therefore $d\le 10^{-2}~m$.

\section{Sedimentation in the boundary layer}
We now study the sedimentation of the different particles in the boundary layer. The purpose is to find out which particles ``fall'' sufficiently 
slowly to be present at the top of the boundary layer, and so to take part in the avalanche.

\subsection{Law relating volumetric concentration and height}
To find the governing law for the volumetric concentration in the boundary layer, 
we use the equations of mass conservation for the mean flow of air and the equation of the  
mean conservation of particles
\begin{equation}
\frac{\partial\overline{u}_{j}}{\partial x_{j}} =0
\end{equation}
\begin{equation}
\frac{\partial\overline{\phi}}{\partial t}+\frac{\partial (\overline{\phi}\overline{u}_{pj})}{\partial x_{j}} =\frac{\partial (D_{t}\frac{\partial\overline{\phi}}{\partial x_{j}})}{\partial x_{j}}\label{7}
\end{equation}
where $\overline{u}_{j}$ are the components of the mean velocity of air, $\overline{u}_{pj}$ the components of the mean velocity of the 
particles, $\overline{\phi}$ the mean volumetric concentration, and $D_{t}$ the turbulent diffusivity that comes from the turbulent 
closure model:
$\overline{\phi 'u_{ip}'}=-D_{t}\frac{\partial\overline{\phi}}{\partial x_{j}}$. $()'$ are turbulent quantities.\\
We suppose that the mean flow is steady and that the velocity and concentration profiles are independent of $x$. With 
the boundary condition that near the ground $\overline{u}_{z} =0$ ($\overline{u}_{z}$ is the velocity perpendicular to the slope), 
we then have $\overline{u}_{z} =0$ in the whole boundary layer. 
Boundary layer theories imply that $D_{t}=\frac{K u^{*} z}{\sigma_{s}}$, where $\sigma_{s}$ is the 
Schmidt number. The Schmidt number is the ratio $\sigma_{s}=\frac{D_{t}}{\nu_{t}}$ between the diffusivity of snow particle and the eddy 
diffusivity of momentum. Some more advanced theories (Bintanja (1998)) take into account that the suspended particles 
modify the turbulence of the air, so 
that the diffusive coefficient $D_{t}$ is modified~: 
\begin{equation}
D_{t}=\frac{K u^{*} z}{\sigma_{s} (1+A R)}
\end{equation}
$$with \ \ R=\frac{-g (\frac{\rho_{p}}{\rho}-1)\frac{\partial\overline{\Phi}}{\partial z}}{\sigma_{s}(\frac{\partial \overline{u}_{x}}{\partial z})^2} \ \ and \ \ A\approx 5-7
$$
where $R$ is the flux ``snowdrift'' Richardson number, $\rho_{p}$ is the density of particles, $\rho$ is the density of air and g is the 
gravitational acceleration. The order of magnitude of the different quantities are the following:\\
$[\overline{\Phi}]\approx 10^{-2}$, $[z]\approx\delta\approx 1~m$, $[\overline{u} ]\approx 50~m.s^{-1}$ and $[\sigma_{s}]\approx 0.5-1$ (see section \ref{pre}). From this we can deduce that $[1+A R]\approx 1$. So within the precision of our study, this modification need not to be taken into 
account.\\ 
To solve the second differential equation, we suppose that the particle volumetric concentration and its gradient vanish infinitely far from the 
boundary layer
\begin{equation}
lim_{z\rightarrow +\infty}\overline{\phi} =0 \ \ \ \  and \ \ \ \  
lim_{z\rightarrow +\infty} \frac{\partial\overline{\phi}}{\partial z} =0\label{8}
\end{equation}
The steadiness of the flow, the independence of the concentration profiles of $x$ and the fact that $\overline{u}_{z} =0$ imply, using equation (\ref{7})
\begin{equation}
\frac{\partial (-\overline{\Phi} v_{r} cos (\alpha )+D_{t}\frac{\partial \overline{\Phi}}{\partial z}}{\partial z}=0
\end{equation}
with $\alpha$ the angle of the slope and $v_{r}=|\overrightarrow{u}_{p} -\overrightarrow{u}|$. From this and with the boundary conditions (see equation \ref{8}) 
we deduce
\begin{equation} 
\overline{\phi} =\overline{\phi}_{1} (\frac{z}{z_{1}})^{-\gamma} \ \ \ \  with \ \ \ \  \gamma =\frac{v_{r} cos(\alpha)\sigma_{s}}{K u^{*}}.
\end{equation}
$\overline{\phi}_{1}$ is the volumetric concentration at $z=z_{1}$.
Let $z_{1}$ be the roughness height, so that $\overline{\phi}_{1}$ is the volumetric concentration of the particles near the ground, i.e., 
the volumetric concentration of the mobilized particles. 
$v_{r}$ is a function of the size of the particles, so the variation in the boundary layer of the volumetric concentration will not be the 
same for different types of particles.

\subsection{Presence of the particles \label{pre}}
\begin{figure}[htbp]
\begin{center}
\epsfig{file=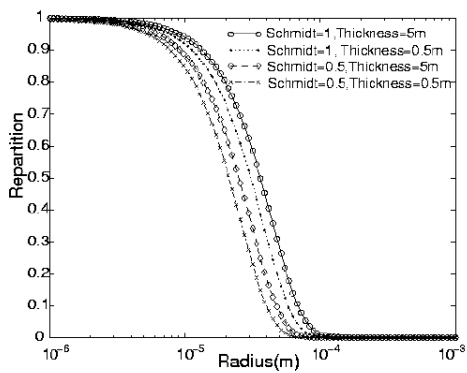,width=7.5cm}
\caption{Repartition for $\sigma_{s}=1$, $\sigma_{s}=0.5$, $\delta =50~cm$ and $\delta =5~m$\label{0.5}}
\end{center}
\end{figure}

To know which particles effectively reach the top of the boundary layer, we study the ratio between the number 
of particles at the top of the boundary layer and the number of them near the ground; to this end we look at the $repartition$ with the radius $a$ 
defined as 
\begin{equation} 
R(a) =\frac{\overline{\phi}}{\overline{\phi}_{1}}=(\frac{z}{z_{1}})^{-\gamma}.
\end{equation}
It defines the concentration of particles by which the body of the avalanche is fed from the boundary layer. To find it 
the relative velocity $v_{r}$ must be determined as a function of $a$. We assimilate it with the sedimentation velocity.
The sedimentation velocity is obtained from the Lagrangian equilibrium state where the gravity force is balanced by the drag force. From this 
equilibrium we obtain
\begin{equation} 
C_{d}Re_{p}^{2}=g(\overline{\rho}_{p}-\overline{\rho})\frac{32 a^3\overline{\rho}}{3\mu^{2}}\label{11}, 
\end{equation}
in which $C_{d}$ is the drag coefficient and $Re_{p}=\frac{2a\rho v_{r}}{\mu}$ is the particle Reynolds number. For spherical particles, as we 
suppose here the snow particles in the flow to be, the drag coefficient can be given as follows, see Ni$\tilde{n}$o and Garcia (1994)
\begin{equation} 
C_{d}=\frac{24}{Re_{p}}(1+0.15\sqrt{Re_{p}}+0.017 Re_{p})-\frac{0.208}
{1+\frac{10^{4}}{\sqrt{Re_{p}}}}\label{2}.
\end{equation}
Another parameter is the Schmidt number ($\sigma_{s}$) which takes different values in different configurations. 
Householder and Goldschmidt (1969)
found that for particles which are denser than the ambient fluid, the Schmidt number is smaller than 1. 
Naaim and Martinez (1995) performed wind tunnel experiments with PVC particles and found a Schmidt 
number of 0.5 to 0.6. We performed calculations both for $\sigma_{s}=1$ and $\sigma_{s}=0.5$. 
The last parameter to fix is $u^{*}$. As before, because we look for the larger range of particle sizes, 
we take $u^{*} =1~m.s^{-1}$. The 
results are reported in Fig~\ref{0.5}. To demonstrate that the dependences on $\sigma_{s}$ and on the thickness of the boundary layer 
are weak we plotted the curves for either $\sigma_{s}=1$ and $\sigma_{s}=0.5$, and either $\delta =50~cm$ and $\delta =5~m$. 
In all cases the 
particles which, in an important number, are still present at the top of the boundary layer, have a size smaller than $2*10^{-4}~m$. 

\section{Conclusion}
The purpose of our study was to estimate the size of the particles which are present in a powder snow avalanche. 
They correspond to the particles that reach the top of the boundary layer, between the ground and the body of the avalanche. Indeed in this way they can take part in the 
processes in
the body of the avalanche.\\
We analyzed the different phenomena which occur in this boundary layer and their influence on the size of the particles. 
Thus, we studied consecutively, the sizes of the particles 
of the snow cover, the 
pick-up of particles by the avalanche and the selection which is done between the particles while 
going up in the boundary layer. With this analyze, we were able to determine the maximum size of the particles which take part in the dynamics of 
a powder snow avalanche. These particles are those which have a size smaller than $200 \mu m$. 
This knowledge will allow to use the results, that have been established in 
diphasic 
flow studies, about the interaction between particles and turbulence. It will also help to improve numerical models which use as 
a parameter the size of the particles of snow. Finally for laboratory experiments, this information is necessary 
 for 
the choice of particles.\\  
Another step, now, is to capture snow particles during the flow 
of a powder snow avalanche, and  measure their size. This will 
help to improve our understanding of the processes and our theoretical model.\\
$\ $\\
{\bf References}\\  
R.~Bintanja.
\newblock {T}he {I}nteraction {B}etween {D}rifting {S}now and {A}tmospheric
  {T}urbulence.
\newblock In {\em Annals of Glaciology}, volume~26, pages 167--173, 1998.\\
J.D. Dent, K.J. Burrell, D.S. Schmidt, M.Y. Louge, E.E. Adams, and T.G.
  Jazbutis.
\newblock {D}ensity, {V}elocity and {F}riction {M}easurements in a {D}ry-{S}now
  {A}valanche.
\newblock In {\em Annals of Glaciology}, volume~26, pages 247--252, 1998.\\
M.K. Householder and V.W. Goldschmidt.
\newblock {T}urbulent {D}iffusion and {S}chmidt {N}umber of {P}articles.
\newblock {\em Journal of the Engineering Mechanics Division, Proceedings of
  the American Society of Civil Engineers}, 95(6):1345--1367, december 1969.\\
J.D. Iversen and B.R. White.
\newblock {S}altation {T}hreshold on {E}arth, {M}ars and {V}enus.
\newblock {\em Sedimentology}, 29:111--119, 1982.\\
M.~Mellor.
\newblock {P}roperties of {S}now.
\newblock {\em Cold Regions Science and Technology}, III(A):1--105, 1964.\\
M.~Naaim and H.~Martinez.
\newblock {E}xperimental and {T}heorical {D}etermination of {C}oncentration
  {P}rofiles and {I}nfluence of {P}article {C}haracteristics in {B}lowing
  {S}now.
\newblock {\em Surveys in Geophysics}, 16:695--710, 1995.\\
K.~Nishimura and Y.~Ito.
\newblock {V}elocity {D}istribution in {S}now {A}valanches.
\newblock {\em Journal of Geophysical Research}, 102(B12):27,297--27,303,
  December 1997.\\
Y.~Ni$\tilde{n}$o and M.~Garc$\acute{\i}$a.
\newblock {G}ravel {S}altation 2.{M}odeling.
\newblock {\em Water Resources Research}, 30(6):1915--1924, June 1994.\\
P.R. Owen.
\newblock {S}altation of {U}niform {G}rains in {A}ir.
\newblock {\em Journal of Fluid Mechanics}, 20(2):225--242, 1964.\\
J.~Qiu, J.~Xu, F.~Jiang, O.~Abe, A.~Sato, Y.~Nohguchi, and T.~Nakamura.
\newblock {S}tudy of {A}valanches in the {T}ianshan {M}ountains, {X}ianjiang,
  {C}hina.
\newblock In {\em Snow Engineering: Recent Advances}, pages 85--90, 1997.\\
K.R. Rasmussen and H.E. Mikkelsen.
\newblock {W}ind {T}unnel {O}bservations of {A}eolian {T}ransport {R}ates.
\newblock {\em Acta Mechanica}, 1:135--144, 1991(supplement).
\end{document}